\documentstyle[editedbook,epsfig,psfig,epsf]{mq}
\def\cm2{cm$^2$ }
\def\se1{s$^{-1}$ }

\def\G1915{GRS~1915+105}
\def\X1550{XTE~J$1550$-$564$}
\def\J1655{GRO~J$1655$-$40$}

\begin{opening}
\title{Low-frequency QPO, Magnetic Flood and the states of GRS~1915+105}
\author{M. Tagger}
\institute{Service d'Astrophysique (CNRS URA 2052), CEA Saclay,
Gif sur Yvette, France.}
\end{opening}

\runningtitle{Magnetic Flood in GRS~1915+105}
\runningauthor{Michel Tagger}

\begin{document}
\vspace{-0.5cm}
\begin{abstract}
{\small  We sum up progress accomplished, since the
last microquasar workshop, on the physics of the Accretion-Ejection
Instability (AEI), and its ability to explain the properties of the low-frequency
QPO of microquasars. These results concern the basic theory of the
instability, its numerical simulation and the resulting modelisation of the
QPO, as well as detailed observations of the QPO properties. They
converge to reinforce the `magnetic flood' scenario, extrapolated from
the AEI to explain the $\sim$ 30 minutes cycles of \G1915. We then
discuss directions in which this scenario might be extended toward a
more global view of the evolution of this source.}
\end{abstract}

%%%%%%%%%%%%%%%%%%%%%%%%%%%%%%%%%%%
\section{Overview}
\label{sec:Overview}
This contribution is first a status report, in which we summarize
recent results on the theory and numerical simulation of the
Accretion-Ejection Instability (AEI), and on detailed observations of
the properties of the low-frequency Quasi-Periodic Oscillation of
microquasars, for which we believe that it provides a convincing
explanation. This progress has been made in complementary
directions (see the contributions of Peggy Varni\`ere and Jer\^ome
Rodriguez, these proceedings):

\begin{itemize}
 \item {\bf Theory:} We have \cite{PV02B} computed what was only
 shown approximately in the original description of the AEI \cite{TP99}, and justifies the `E' in its name: its unique ability to send upward, as Alfv\'en waves propagating in the corona, the accretion energy  and angular momentum it extracts from the disk. We find that this mechanism is highly efficient, so that a large fraction of the accretion energy can be converted into Alfv\'en waves. Future work should show how this energy can be deposited in the corona to energize a wind or jet.
\item {\bf Numerical simulation:} %
We have improved the numerical simulations presented by S.~Caunt \cite{CT00} at the last workshop.  A simple model of heating and thickening of the disk  at the spiral shock formed by  the instability allows us to start
producing\footnote{in collaboration with M. Muno(MIT)} synthetic light curves, which can be compared with the observed ones. Preliminary results show in particular that the AEI can reproduce the high rms amplitude of the QPO.

\item{\bf Observation:} We have now published in final form the results presented at the last workshop, comparing  theory and observation of relativistic effects when the inner disk edge approaches the last stable orbit, in particular in \J1655 and \G1915 \cite{JR02A}-\cite{PV02A}. Although the observational  evidence is fragile  and can only be taken as an indication, an additional hint is now provided by observations of \X1550 \cite{JR02B}.
 \item {\bf Energy spectrum of the QPO:} It is commonly observed that  the QPO is best correlated with the properties of the disk, although  it affects more strongly the coronal emission, {\em i.e.} the power-law tail. However, in the same observations of \X1550, we find  that the modulated emission has a different energy spectrum than this power-law. This is consistent with the presence of a hot point  in the disk {\em i.e.},  in our interpretation, the spiral shock.

\end{itemize}
Thus these results converge to confirm the expected properties of the AEI and its ability to explain the main characteristics of the QPO. These expectations were the basis of the `Magnetic Flood' scenario\cite{Flood99}, which starts from the identification of the QPO with the AEI, and extrapolates to give a tentative explanation for the $\sim$30 minutes cycles of \G1915.  In this scenario the cycles are determined by the processing of the poloidal (vertical) magnetic flux advected with the gas in the disk.  We will show below how we have found a positive indication in favor of this scenario, in the detailed analysis of one of these cycles.  

We will then turn to unpublished work, presented in the undergraduate thesis of Fitzgibbon (1999) with E. Morgan and R. Remillard at MIT: defining states analogous to the ones of Belloni {\em et al.}\cite{BELLONI00}, they show a striking regularity in the succession of these states: the source does not err between them, but repeatedly follows a well-defined sequence between them. We will briefly discuss how the  Magnetic Flood scenario could be extrapolated again to explain  this regularity.

\section{The Magnetic Flood Scenario}
\label{sec:Flood}
Presented previously \cite{Flood99} and at the last workshop, this starts from  the identification of the AEI as the origin of the QPO, and follows a simple line of inferences which could explain the 30 mn cycles of \G1915. Its first step is that the apparition of the QPO, when the disk transitions from the high-soft to the low-hard state, must correspond to the crossing of a stability threshold. In our scenario the QPO appears when the magnetic pressure in the inner region of the disk builds up, reaching  equipartition with the gas pressure; this is known to suppress the Magneto-Rotational turbulence (causing accretion in weakly magnetized disks), and conversely it is the most unstable situation for the AEI. 

As a consequence the disk cools down, since the AEI transports the accretion energy outward and upward, in the form of wave flux, rather than depositing it locally to heat the gas (as assumed in the $\alpha$-disk model). This explains that the disk transitions to the low-hard state, characterized by weak disk emission and  dominant coronal one, and by the presence of a strong QPO. The disk moves outward and then back in, as it is viscously refilled from the outer regions; the low state ends when the inner disk edge reaches the Last Stable Orbit, allowing its magnetic flux to reconnect with that trapped in the vicinity of the black hole (this may be the same force-free magnetic structure associated with the Blandford-Znajek mechanism).  The reconnection, seen as the `spike'  ending the low-hard state (see the `Magnetic Bomb' model presented by S. Eikenberry, this workshop) causes the ejection of the relativistic plasmoids later seen in IR and radio, and destroys magnetic flux: this allows a return to a lower magnetization state where the magneto-rotational instability dominates again  the accretion process. 

Our detailed analysis of two such cycles was mainly motivated by a comparison of observations with the theory of relativistic effects on the AEI, mentioned in the previous section. However we also found that the QPO appeared shortly  {\em before} the transition to the low state. Thus, if there is a causal relation  between the low state and the QPO, it is not the one usually assumed:  the QPO is not a property of the low state, but might well be what  causes it, as predicted by our scenario.  
\begin{figure}[htb]
%\centering
%\vspace{0cm}
\vspace{-1cm}
\hspace*{-1cm}
\epsfig{file=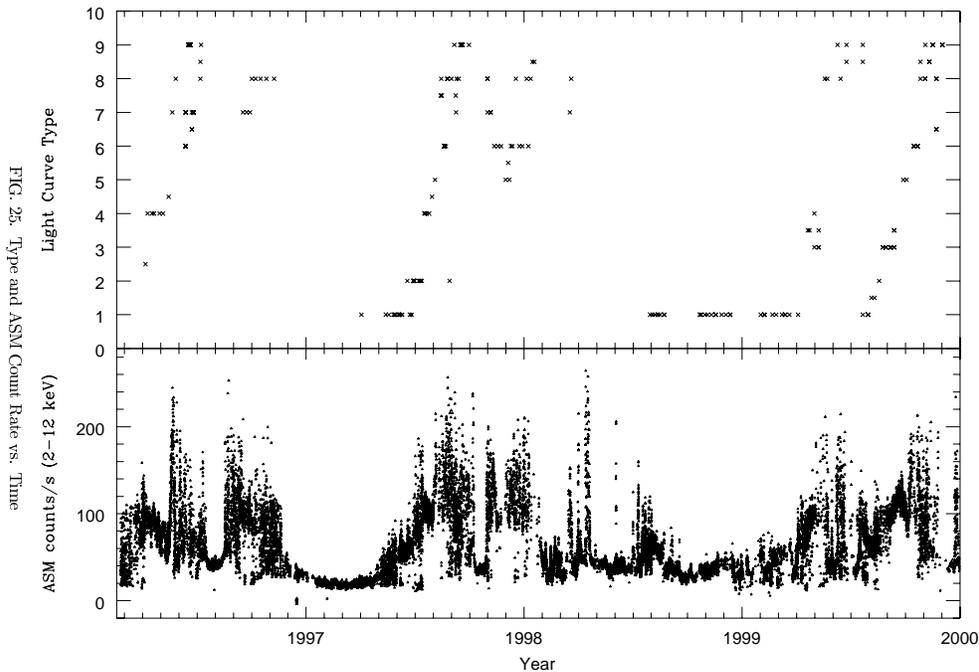,width=12cm, angle=-90}
\vspace{-1.5cm}
\caption{{\em Bottom:}ASM count of \G1915 from 1996 to 2000. 
{\em Top:} State of the source in Fitzgibbon's categories.}
\label{fig:Fitz}
\end{figure}
This scenario is compatible with all  the observational facts; we use it as a guide in our ongoing work of theory, simulation and observation; this new fact lets us hope that it may become a true model of the evolution of this source, from which much could be learnt: concerning other states of \G1915, its long-term behavior, and by extension the physics  of accretion in black-hole binaries. 

\section{\ldots and beyond?}
\label{beyond}
One way to progress in this direction might be to turn to the results of Fitzgibbon, reproduced in Figure \ref{fig:Fitz}: he defined 9 states of the source, similar to the 12 states of Belloni {\em et al.} (a tenth state would be Belloni's state $\chi$, and is not shown). Figure~\ref{fig:Fitz} is his main result: he found that the source goes from one state to the other in a well-defined order, following the {\em history} of its X-ray luminosity, rather than simply the luminosity itself with which it is only loosely correlated: as plateau states are followed by flares, the source transitions from Fitzgibbon's states 1 to 8 and 9 (Belloni's states $\beta$ and $\vartheta$, subject to the 30 mn cycles). They are followed by a drop back to the plateau state, until the next flare.

This result, although unpublished, may be of great importance as it  probably contains a clue to the additional parameter, besides the accretion rate, which is suspected to rule the behavior of \G1915. This must have to do with the  history of the source, which cannot return to the plateau state before it has gone through the 30 mn cycles, and before that through the succession of the different states. We suggest that this parameter could be the poloidal flux advected in the disk: in the Magnetic Flood scenario, it is its piling up in the inner region of the disk which causes the cycles.

Extrapolating further this scenario, we thus get to the following working hypothesis: although one might think of other  possibilities, reconnection is the best candidate to explain the spike and the ejection of the plasmoid; but this requires that the magnetic fields in the disk and in the `hole' inside the Last Stable Orbit have opposite polarities. We thus assume that the poloidal field trapped near the black hole is sometimes parallel and sometimes anti-parallel to the one frozen in the disk (in the solar wind--earth magnetosphere system, the same dichotomy allows reconnection in the anti-parallel case, while in the parallel case more complex physics is involved). On long time scales the poloidal flux advected with the gas can change polarity; this may be due to the dynamo in the companion star, or in the disk itself. In a parallel configuration accretion causes the flux trapped in the vicinity of the black hole to gradually build up until, after the next change of polarity in the disk, it can start to be destroyed by reconnection. This (the anti-parallel configuration) would correspond to the flares, and the cycles in Fitzgibbon's states 8 and 9 would occur when the trapped flux has dropped and will soon also change sign: in this case the return to the plateau state corresponds to the return to the parallel fields configuration, and the beginning of the flares to the next field reversal in the disk and the onset of an anti-parallel phase. 

At this stage this is only an extrapolation of the Magnetic Flood scenario, which is itself an extrapolation of the identification of the AEI as the source of the QPO - although we believe that our recent results have increased its credibility. We note, however, that it  would nicely explain that the plateau and the flares seem to occupy roughly comparable durations over the period studied by Fitzgibbon, since in order to annihilate the magnetic  flux built up during one episode the disk has to accrete a comparable amount of opposite flux. We hope that analyzing other states of \G1915 may help in refining this hypothesis, leading to a more global view of accretion physics in this and other sources.

\section*{Acknowledgments}
The author thanks Peggy Varni\`ere and Jer\^ome Rodriguez for the work summarized in this paper, and for their contribution in discussing Fitzgibbon's results.

\end{document}